\documentclass[UTF-8]{article}
\usepackage{graphicx}
\title{Aiming to Detect a malware of GSM frequency}
\author{{\small Weijun ZHU and \small Kai NIE (equal contribution)}\\
{\footnotesize School of Information Engineering, Zhengzhou University, Zhengzhou, 450001, China}\\
}

\begin{document}

\maketitle

\begin{abstract}
In order to find a specific type of malware/attack which is related to GSM frequency, we propose an algorithm according to the most essential characteristics of this malware. At first, detect whether or not there exists a specific thread in the memory. And then, the generated binary strings will be tried to be matched with the one in the target computer. At last, determine whether this threat occurs or not. Furthermore, we study the effective of the new method via some simulations.
\end{abstract}
\section{Introduction}
Intrusion detection (ID) is an important network security technique, which detects whether there exist intrusion attacks in the network through the use of some technical means. If an attack is detected, the ID alerts the user. In fact, the attack and the defense is always a pair of contradiction in the network security. Along with the development of network defense techniques including intrusion detection, the network attack technique is also escalating. On the one hand, the variant of the known attacks makes the traditional intrusion detection system more difficult to detect. On the other hand, some network attacks have occured  in the last three years, which was viewed as impossible by the traditional secure opinion. For exanmple, an air-gapped networks is no longer secure under the network attack. And there are a number of techniques that allow an attacker to theft information from an air-gapped computer at present.

Since 2013, some new intrusion techniques are emerging. And these types of attacks have been developed to theft data from a air-gapped computer by using the thermal emission [1], electromagnetic radiation [2], ultrasonic [3][4], or USB devices [5]. Traditional intrusion detection methods can do nothing about these attacks, including GSMem, a malware for air-gapped computers. In this paper. we aim to deal with this problem.

\section{The GSMem Attack Model}
Air-Gapped Networks are physically and logically isolated from the public Internet, and this section describes "GSMem" attacks that exfiltrate data from air-gapped networks. "GSMem" is a bifurcated malware, which consists of two parts, part of the program is installed on the desktop computer to achieve the function of the signal transmitter, the other part of the program running in the baseband mobile phone to achieve the function of the signal receiver. The desktop computer generates the electromagnetic radiation (EMR) of the handset baseband-compatible GSM band by calling the memory-related CPU instructions [6][7][8][9][10]. By increasing the amplitude of the electromagnetic waves using the multi-channel memory architecture function, thereby modulating the (binary) information desired to be stepped into the amplified electromagnetic signals. A standard GSM mobile phone by modifying the baseband processor to receive and decode these amplified emission of electromagnetic signals, resulting in data containing user privacy. The following describes the  principle of  the transmitter and the receiver in "GSMem" attack.

the core principles and key steps of GSMem attacks are shown in Figure 1.
\begin{figure}
	\centering
	\scalebox{0.7}{\includegraphics{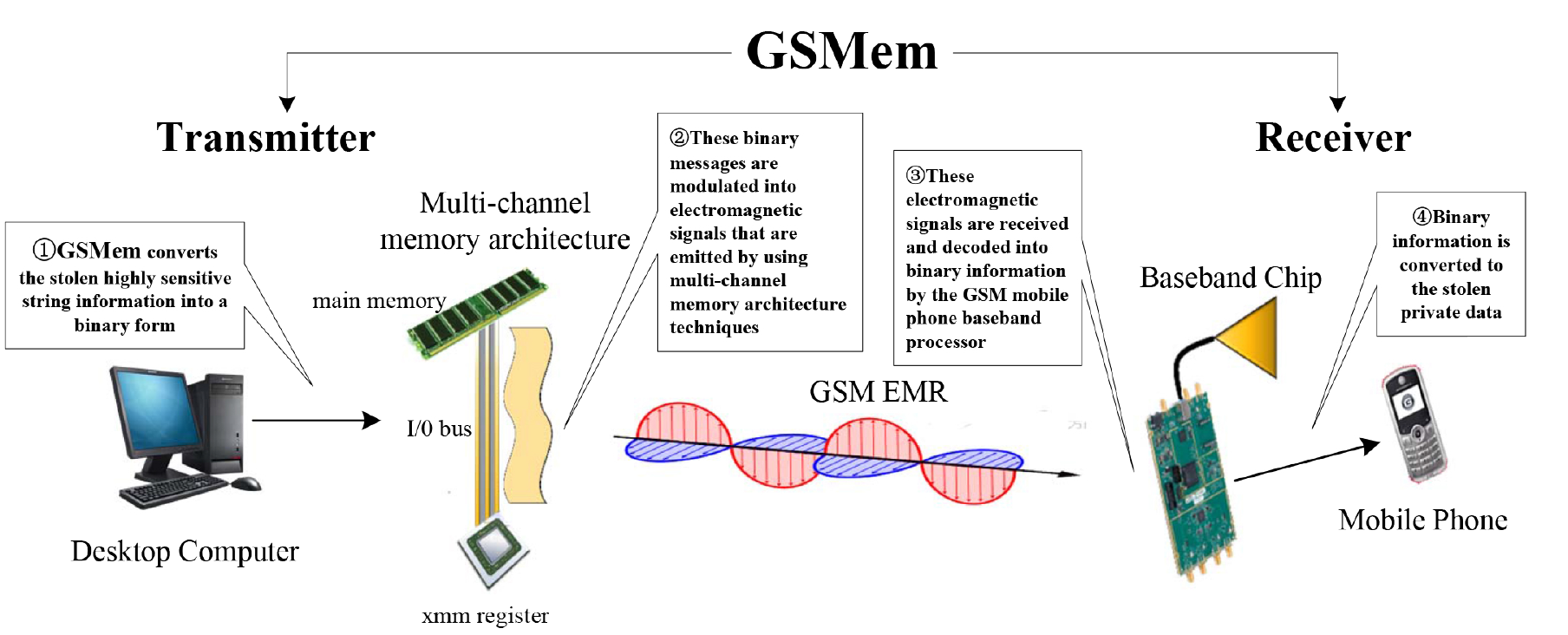}}
	\caption{ GSMem attack schematic}
\end{figure}
\section{ An Algorithm for detecting GSMem}
The transmitter in the GSMem attack model that operates on a desktop computer, and the transmitting program has a small memory and CPU footprint, making the activities of the transmitter easier to hide and difficult to be found.

The main reasons are:\textcircled{1}In terms of memory consumption, the program consumes merely 4K of memory allocated on the heap.\textcircled{2}In terms of CPU intake, the transmitter runs on a single, independent thread.\textcircled{3}At the OS level, the transmitting process can be executed with no elevated privileges (e.g., root or admin).\textcircled{4}Finally, the code consists of bare CPU instructions, avoiding API calls to escape certain malware scanners.

In short, the transmission code evades common security mechanisms such as API monitoring and resource tracing, making it hard to detect.

However, the transmission program is difficult to detect does not mean that can not be detected. According to the most essential characteristic of the transmitter in the GSMem attack model, we propose the following algorithm for GSMem attack detection from the call of MOVNTDQ instruction.
\begin{figure}
	\centering
	\scalebox{0.75}{\includegraphics{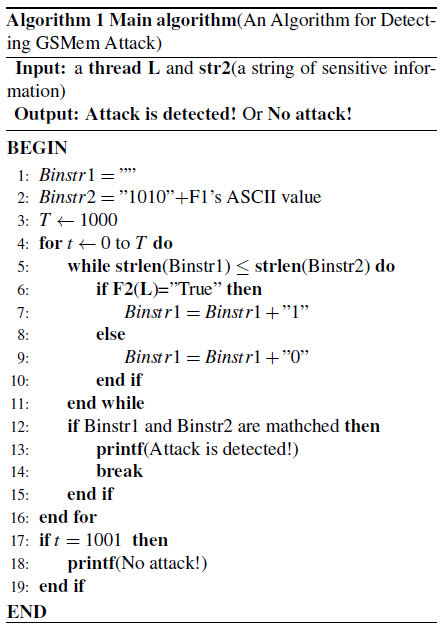}}
	\caption{Algorithm 1}
\end{figure}
\begin{figure}
	\centering
	\scalebox{0.75}{\includegraphics{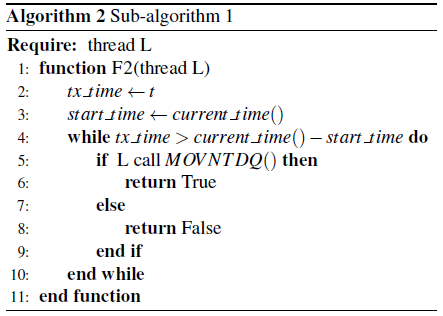}}
	\caption{Algorithm 2}
\end{figure}
The new algorithm consists of a main algorithm and a sub-algorithm. Since the GSMem is a bifurcated malware and the transmitter is installed on a desktop computer, the main algorithm starts with the detection of the transmitter in the GSMem attack model.

The critical to the GSMem attack model is that the transmitter uses the MOVNTDQ instruction to generate the electromagnetic signal to transmit data (ie, "1" and "0" in the amplitude of the electromagnetic waves emissions) by the attack principle. According to the nature of this principle, the existence of the thread can call MOVNTDQ start detection. First, the new algorithm detects whether there is such a thread in memory, which is done by sub-algorithm 1. Of course, the new algorithm must exclude the normal use of MOVNTDQ instruction thread. Next, the new algorithm divides the time "T" one by one into 1ms to 1000ms according to the "B-ASK" modulation principle of the transmitter in the GSMem attack model. The binary string generated by a single program loop in the algorithm (including the preamble sequence "1010") matches the binary string of sensitive information. Once the two binary strings match, that is found GSMem attack, and the attacker is sending high-level sensitive information, then the algorithm immediately alarm! 

\section{Simulation Experiments}
In this section, Matlab is employed to conduct our simulations. And the process of the algorithm is simulated in this platform.
\subsection{Experiment Setup}
First of all, we start from the basic principle of algorithm detection, simulation thread calls MOVNTDQ, simulation of the main algorithm and sub-algorithm implementation, the specific experimental steps are as follows:\\
(1) Input the string of sensitive information, and convert the string into binary form.\\
(2) Electromagnetic waves sampling. We according to the literature [3] to provide normal use of desktop computers and GSMem attacks in the case of amplitude-frequency diagram, sampling the frequency and amplitude of the electromagnetic waves emissions from the normal and the GSMem attack, this provides accurate raw data for this experiment. More precisely, it is a lot of raw data about normal behavior and attack behavior.\\
(3) The simulation thread periodically calls MOVNTDQ instruction. We simulate the electromagnetic waves emitted by the computer at the moment. If the frequency and amplitude of the emitted electromagnetic waves (sampled from (2)) fall within the range of the anomalous electromagnetic waves (sampled from (2)) emanating from the transmitter in the GSMem attack model at the current moment. The thread is considered to have invoked the MOVNTDQ instruction and continues the detect (proceed to (4)).\\
(4) We use the string matching principle of sub-algorithm 3 in the new algorithm to detect whether sensitive information is exfiltrated in the thread regular use of MOVNTDQ instruction. Specifically, according to the time slice one by one to generate the binary string in the algorithm and the binary string of sensitive information in the step (1) to match, if the match is successful, we think that find an attack, immediately alarm!
\subsection{Experimental results}
\begin{figure}
	\raggedleft
	\scalebox{0.4}{\includegraphics{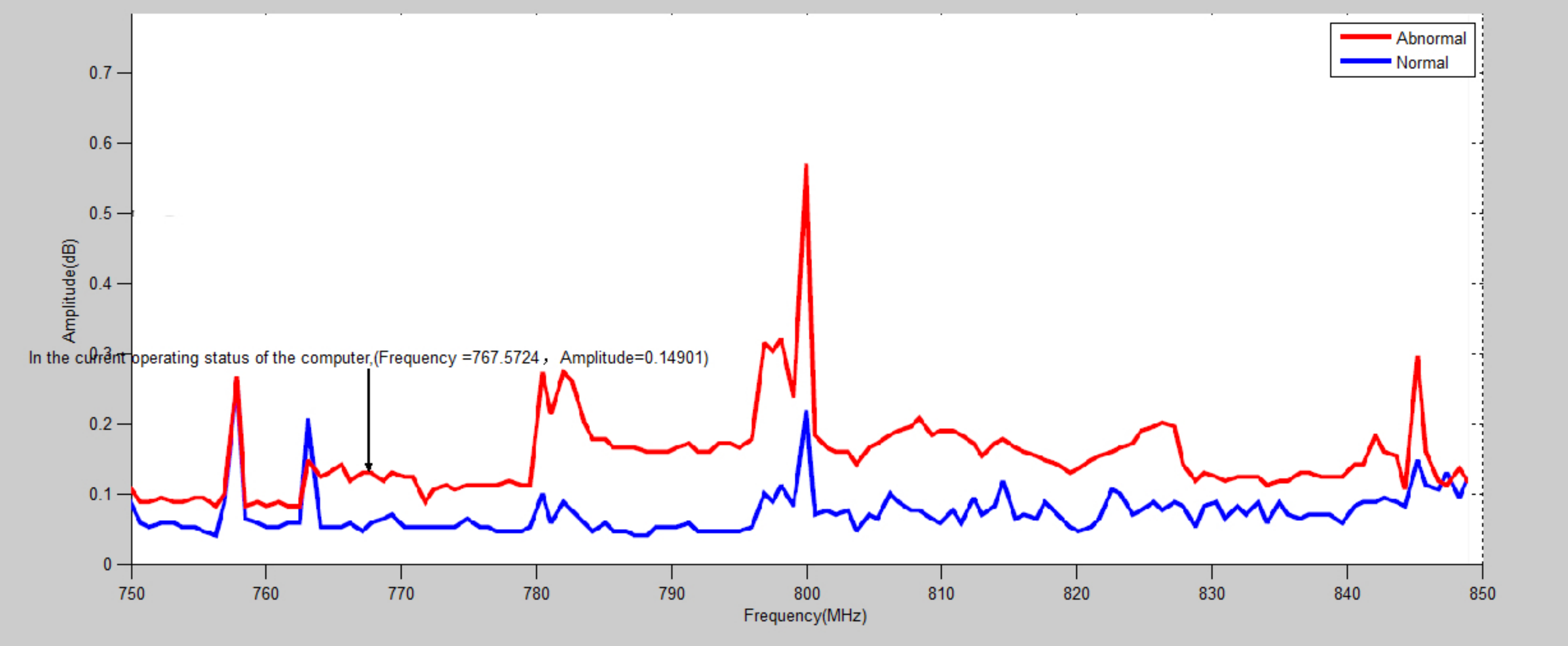}}
	\caption{abnormal electromagnetic waves}
\end{figure}
\begin{figure}
	\centering
	\scalebox{0.9}{\includegraphics{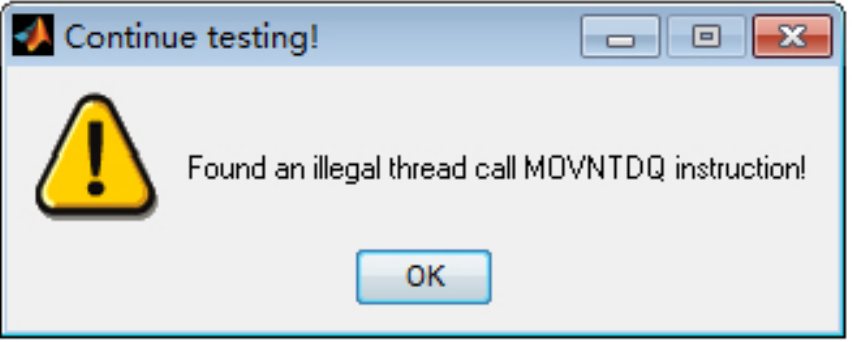}}
	\caption{found thread calls the MOVNTDQ instruction}
\end{figure}
\begin{figure}
	\centering
	\scalebox{0.9}{\includegraphics{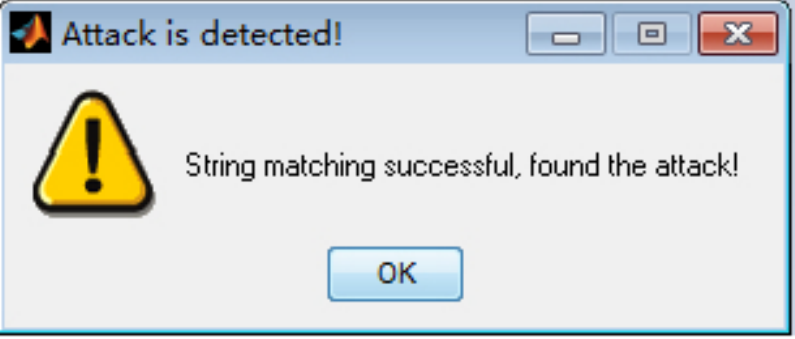}}
	\caption{Successful string matching}
\end{figure}
If a computer is currently running radiated electromagnetic waves in the GSMem emitter radiated from the range of abnormal electromagnetic waves, the results shown in Figure 4. The figure shows the frequency and amplitude of the anomalous electromagnetic waves, which means that there is a possibility that a suspect attack uses this frequency and amplitude to transmit information. If you further detect illegal call MOVNTDQ instruction, the results shown in Figure 5, means that the possibility of suspected attacks increased. Thereafter, if the binary string matches successfully, the result is shown in Figure 6, at this time it means that find an attack!

\section{Conclusions}
The main result of this paper is Algorithm 1, which aims to detect GSMem. This is the target of the new approach.
\section*{Acknowledgements}
This work has been supported by the Natural Science Foundation of China under Grant No.U1204608 as well as China Postdoctoral Science Foundation under Grant No.2012M511588 and No.2015M572120.

\end{document}